\documentclass[aps]{revtex4}
\usepackage{graphicx}
\usepackage{dcolumn}
\usepackage{bm}
\usepackage{rotate}
\usepackage{epsfig}
\newcommand \be  {\begin{equation}}
\newcommand \bea {\begin{eqnarray} \nonumber }
\newcommand \ee  {\end{equation}}
\newcommand \eea {\end{eqnarray}}
\begin{document}
\title{Trend followers lose more often than they gain}
\author{Marc Potters$^*$, Jean-Philippe Bouchaud$^{*,+}$}
\email{marc.potters@cfm.fr,jean-philippe.bouchaud@cfm.fr}
\affiliation{
$^*$ Science \& Finance, Capital Fund Management, 6-8 Bd
Haussmann, 75009 Paris, France.\\
$^{+}$ Service de Physique de l'{\'E}tat Condens{\'e},
Orme des Merisiers,
CEA Saclay, 91191 Gif sur Yvette Cedex, France.
}
\date{\today}

\begin{abstract}
We solve exactly a simple model of trend following strategy, and obtain the analytical shape
 of the profit per trade distribution. This distribution is non trivial and has an option
like, asymmetric structure. The degree of asymmetry depends 
continuously on the parameters of the strategy and on the volatility of the traded asset. 
While the average gain per trade 
is always exactly zero, the fraction $f$ of winning trades decreases from $f=1/2$ for 
small volatility to $f=0$ for 
high volatility, showing that this winning probability 
does not give any information on the reliability of the strategy 
but is indicative of the trading style.
\end{abstract}

\maketitle

A question often asked by investors to fund managers, besides the average return of their strategies, is: ``What 
is your fraction of winning trades?" Implicitly, they expect the answer to be larger than $50 \%$ , as this would 
indicate that the fund manager is more frequently right than wrong, and {\it therefore} trustworthy. 
We want to show 
in this paper that this fraction is in fact meaningless. It depends entirely on the trading style of the manager, 
and tells very little about the consistency of his returns. It is clear that one can make money on average even if the 
fraction $f$ of winning trades is low, provided the average gain per winning trade $\cal G$ exceeds the average loss per 
losing trade, $\cal L$. The condition is, clearly:
\be
f {\cal G} > (1-f) {\cal L}.
\ee
Since asset prices are very close to being pure random walks, any statistical signal exploited by systematic 
traders has an extremely small signal to noise ratio. The average profit {\it per trade} for any hedge fund is
bound to be very small, which means that the above inequality is close to an equality. If the typical holding 
period of winning trades is $T_G$ and that of losing trades $T_L$, one expects, for a random walk
of volatility $\sigma$:
\be
{\cal G} \sim \sigma \sqrt{T_G}, \qquad {\cal L} \sim \sigma \sqrt{T_L}.
\ee
Therefore, the fraction of winning trades is in fact a measure of the ratio of the holding periods of winning trades
to that of losing trades:
\be
\frac{T_G}{T_L} \approx \left(\frac{1-f}{f}\right)^2.
\ee
For example, a $40 \%$ fraction of winning trades merely indicates that (unless the manager is really lousy) the typical 
holding period of winning trades is $\sim 2$ times that of losing trades. This, in turn, means that the manager is
probably mostly trend following, since by definition a trend following strategy stays in position when the move
is favorable, but closes it in case of adverse moves. Hence, conditioned to a winning trade, the holding period
is clearly longer. The opposite would be true for a contrarian strategy. Let us illustrate this 
general idea by two simple models. The first one is completely trivial and not very interesting besides driving our
point home. The second model is much richer; it can be solved exactly using quite interesting methods from the 
theory of random walks and leads to a very non trivial distribution of profits and losses. Besides its intrinsic 
interest, the model elegantly illustrates various useful methods in quantitative finance, and could easily be used 
as a basis for a series of introductory lectures in mathematical finance. The outcome of our 
calculations is that although the daily P\&L of the strategy is trivial and reflects the statistics
of the underlying, the P\&L of a {\it given trade} has an asymmetric, option-like structure!   

The first model is the following: suppose that the price, at each time step, can only move up $+1$ 
or down $-1$ 
with probability $1/2$. 
The trend following strategy is to buy (sell) whenever the last move was $+1$ ($-1$) and the previous position was
flat, stay in position if the last move is in the same direction, and close the position as soon as the move is adverse.
Conditioned to an initial buy signal, the probability that the position is closed a time $n$ later is clearly:
\be
p_n=(1/2)^n.
\ee
If $n=1$, the trade loses $1$; whereas if $n > 1$, the trade gains $n-2$. Therefore, the average gain is, obviously:
\be
\sum_{n=1}^\infty (n-2) \, (1/2)^n  = 0,
\ee
as it should, whereas the probability to win is $1/4$, the average gain per winning trade $\cal G$ is $2$ and 
the average loss per losing trade $\cal L$ is $1$. The average holding period for winning trades is:
\be
T_G = 4 \sum_{n=3}^\infty n (1/2)^n = \frac92.
\ee

Let us now turn to our second, arguably more interesting model, where the log-price $P(t)$ is assumed to 
be a continuous time Brownian motion. This model has well known deficiencies: the main drawbacks of the model
are
 the absence of jumps, volatility fluctuations, etc. that make real prices strongly non-Gaussian and 
distributions fat-tailed \cite{Book}. However, for the purpose of illustration, and also because the continuous time Brownian motion
is still the standard model in theoretical finance, we will work with this model, which turns out to be exactly
soluble. We assume that on the time scale of interest, the price is driftless, and write:
\be
\frac{dP}{P} = \sigma dW(t),
\ee
where $dW(t)$ is the Brownian noise and $\sigma$ the volatility. From these log-returns, trend followers form the 
following ``trend indicator" $\phi(t)$, obtained as an exponential moving average of past returns:
\be\label{dphi}
\phi(t) = \sigma \int_{-\infty}^t e^{-(t-t')/\tau} dW(t'),
\ee
where $\tau$ is the time scale over which they estimate the trend. Large positive $\phi(t)$ means a solid trend up over the
last $\tau$ days. Alternatively, $\phi(t)$ can be written as the solution of the following stochastic differential 
equation:
\be
d\phi = -\frac1\tau \phi \, dt +  \sigma dW(t).
\ee
The strategy of our trend-follower is then as follows: from being initially flat, he buys $+1$ when $\phi$ reaches the
value $\Phi$ (assuming this is the first thing that happens) and stays long until $\phi$ hits the value $-\Phi$, at which 
point he sells back and takes the opposite position $-1$, and so on. An alternative model is to close
the position when $\phi$ reaches $0$ and remain flat until a new trend signal occurs. 
The profit $G$ associated with a trade is the total
return during the period between the opening of the trade, at time $t_o$ ($\phi(t_o)= \pm \Phi$) and the closing of the 
same trade, at time $t_c$ ($\phi(t_c)=\mp \Phi$). More precisely, assuming that he always keeps a constant investment
level of $1$ dollar and neglecting transaction costs,
\be
G = \int_{t_o}^{t_c} \sigma dW(t).
\ee
We are primarily interested in the profit and loss distribution of these trend following trades, that we will denote
$Q(G)$. A more complete characterization of the trading strategy would require the joint distribution $Q(G,T)$ of $G$ 
on the one hand, and of the time to complete a trade $T={t_c}-{t_o}$ on the other; this quantity is discussed in the
Appendix. Obviously, $G$ and $\phi$ evolve in a correlated way, since both are driven by the noise term $dW$.
For definiteness, we will consider below the case of a buy trade initiated when $\phi= + \Phi$; since we assume the
price process to be symmetric, the profit distribution of a sell trade is identical. Now, the trick to solve the problem 
at hand is to introduce the conditional distribution $P(g|\phi,t)$ that at time $t$, 
knowing that the trend indicator in $\phi$, the profit still to be earned between $t$ and $t_c$ is $g$. This 
distribution is found to obey the following {\it backward} 
Fokker-Planck equation:
\be
\frac{\partial P(g|\phi,t)}{\partial t} = - \frac{\phi}{\tau} \frac{\partial P(g|\phi,t)}{\partial \phi} +
\frac{\sigma^2}{2} \left[\frac{\partial^2 P(g|\phi,t)}{\partial \phi^2} - 2 
\frac{\partial^2 P(g|\phi,t)}{\partial \phi \partial g} + \frac{\partial^2 P(g|\phi,t)}{\partial g^2}\right].
\ee
However, since $\phi$ is a Markovian process, it is clear that the history is irrelevant and at any time $t$, the
distribution of profit still to be made only depends on how far we are from reaching $\phi = - \Phi$, that is, $P(g|\phi,t)$
depends on $\phi$ but not on $t$. Therefore, one finds the following partial differential equation:
\be\label{central}
- {\phi} \frac{\partial P(g|\phi)}{\partial \phi} +
\frac{\sigma^2 \tau}{2} \left[\frac{\partial^2 P(g|\phi)}{\partial \phi^2} - 2 
\frac{\partial^2 P(g|\phi)}{\partial \phi \partial g} + \frac{\partial^2 P(g|\phi)}{\partial g^2}\right]=0.
\ee
Eq. (\ref{central}) has to be supplemented with boundary conditions: obviously when $\phi=-\Phi$ the 
yet to be made profit must be zero, imposing:
\be
P(g|\phi=-\Phi)=\delta(g).
\ee
The final quantity of interest is the profit to be made when entering the trade, i.e:
\be
Q(G)=P(g=G|\phi=+\Phi).
\ee
We now proceed to solve Eq.  (\ref{central}). First, it is clear that all the results can only depend on the ratio 
$\Phi/\sigma \sqrt{\tau}$, i.e. on the width of the trend following `channel' $\Phi$ measured in units of the typical 
price changes over the memory time $\tau$, that is, the order of magnitude of the expected gains of the trend following 
strategy. One expects in particular that in the limit $\Phi/\sigma \sqrt{\tau} \to \infty$, the distribution of 
gains will become Gaussian, since the time needed to reach the edge of the channel is then much larger than the memory 
time of the process. We will from now on measure $\Phi$ and $G$ in units of $\sigma \sqrt{\tau}$, and therefore set 
 $\sigma^2 \tau = 1$ hereafter. Now, Fourier transforming $P(g|\phi)$ with respect to $g$:
\be
P(g|\phi)= \int \frac{{\rm d}\lambda}{2 \pi} \, e^{i \lambda g} \Psi_\lambda(\phi),
\ee
one obtains an ordinary differential equation for $\Psi_\lambda(\phi)$:
\be\label{ode}
\frac{\partial^2 \Psi_\lambda(\phi)}{\partial \phi^2} - 2 (\phi + i \lambda) 
\frac{\partial \Psi_\lambda(\phi)}{\partial \phi} - \lambda^2 \Psi_\lambda(\phi) = 0.
\ee
This is known as the Kummer equation (or, after a simple transformation, as the Weber equation) 
\cite{Grad,HarmOsc}. 
The general solution 
is the sum of two confluent Hypergeometric functions $_1F_1$, with coefficients that are determined by two 
boundary conditions. We already know that the boundary condition at $-\Phi$ should be $\Psi_\lambda(-\Phi)=1$, 
$\forall \lambda$. The second boundary condition turns out to be that $\Psi_\lambda(\phi)$ should be well behaved for $\phi \to \infty$, i.e.
not grow exponentially with $\phi$. A way to be convinced and get some intuition on the solution is to 
expand $\Psi_\lambda(\phi)$ for small $\lambda$ as:
\be
\Psi_\lambda(\phi) = \psi_0(\phi)+i \lambda \psi_1(\phi) - \frac{\lambda^2}{2} \psi_2(\phi) + ...
\ee
Plugging this into Eq. (\ref{ode}), one finds:
\be
\psi_0(\phi) \equiv 1; \qquad \psi_1(\phi) \equiv 0; \qquad \psi_2'' - 2 \psi_2'= - 2.
\ee
The first two results are expected and simply mean that $P(g|\phi)$ is normalized for all $\phi$, and that the
average gain is identically zero, as must indeed be the case of any strategy betting on a random walk. The last equation
is more interesting; the only reasonable solution of this equation is: 
\be
\psi_2(\phi) = 2 \int_{-\Phi}^\phi {\rm d}u \, e^{u^2} \int_u^\infty {\rm d}v \, e^{-v^2},
\ee
which for large $\phi$ behaves as $\ln \phi$. This is indeed expected: if the trade did open 
when $\phi$ hits a very large value instead of at $+\Phi$, the time needed for $\phi$ to come back to values of order $\Phi$ can
be obtained by solving Eq. (\ref{dphi}) without the noise term, giving $T \sim \tau \ln \phi$. The total gain is the 
sum of $\sim T/\tau$ random contributions, its variance is thus expected to be $\sim \ln \phi$. In fact, in the
limit $\phi \to \infty$, the distribution of gains indeed becomes exactly Gaussian, as can be seen by writing:
\be
\Psi_\lambda(\phi) = e^{-\frac{\lambda^2}{2} Z_\lambda(\phi)}.
\ee
The corresponding {\sc ode} for $Z_\lambda(\phi)$ reads:
\be\label{ZZ}
\frac{\lambda^2}{2} Z'^2 + i \lambda Z' - Z'' + 2 \phi Z' - 2 = 0,
\ee
from which one immediately finds that for $\phi \to \infty$, $Z \sim \ln \phi$ independently of $\lambda$. Therefore,
in that limit, the characteristic function $\Psi_\lambda(\phi)$ indeed becomes Gaussian (in $\lambda$ and thus in $g$). 
The above equation on $Z$ will be useful below to extract the large $\lambda$ behaviour of $\Psi_\lambda$. 
The conclusion of this analysis is that the large $\phi$ behaviour of $\Psi_\lambda(\phi)$ is a decreasing power-law:
\be
\Psi_\lambda(\phi) \sim \phi^{-\lambda^2/2}.
\ee
This gives us our second boundary condition. The correct solution of our problem can then be written as:
\be
\Psi_\lambda(\phi) = \frac{W_\lambda(\phi)}{W_\lambda(-\Phi)},
\ee
with $W$ the following combination of hypergeometric functions:\footnote{Note that we cannot write $W$ as the Kummer function of the second kind $U$ 
because as we follow the solution from $-\Phi$ to $+\Phi$ we run into a branch cut of $U$ at $\phi=0$ when the third 
argument falls on the negative real axis. Eq (24), on the other hand, does not have a branch cut at $\phi=0$.}
\be\label{W}
W_\lambda(\phi)= \, _1F_1\left(\frac{\lambda^2}{4},\frac{1}{2},(\phi + i \lambda)^2\right) - 
2 \frac{\Gamma(\frac{\lambda^2}{4}+\frac{1}{2})}
{\Gamma(\frac{\lambda^2}{4})} (\phi + i \lambda)\, _1F_1\left(\frac{\lambda^2}{4}+\frac{1}{2},\frac{3}{2},
(\phi + i \lambda)^2\right),
\ee
related to the so-called the Weber function \cite{HarmOsc}. 
One can check, using the known asymptotic behaviour of the hypergeometric functions, that this particular 
combination indeed decays as $\phi^{-\lambda^2/2}$ for large $\phi$. 

From these expressions, one can reconstruct the whole distribution $Q(G)$, that we now describe. As already mentioned
above, in the limit $\Phi \to \infty$ the distribution becomes Gaussian. As $\Phi$ decreases, the distribution becomes
more an more positively skewed: the fraction of winning trades decreases, but the average gain per winning trade becomes 
larger. This is illustrated in Fig. 1 where we plot $Q(G)$ for the intermediate case $\Phi=1$. It is clear that the 
most likely profit is negative; the probability to lose is in that case $1-f \approx 0.635$. The distribution can be characterized
further by studying its asymptotic tails for $G \to \pm \infty$. This can be done by observing that  $\Psi_\lambda(\phi)$
has poles for $\lambda$ imaginary, corresponding to zeros of $W_\lambda(-\Phi)$. For $\Phi=1$, we find that the zeros
closest to $\lambda=0$ are $\lambda_+=0.432i$ and $\lambda_-=-5.058 i$, translating into the following large $|G|$ 
behaviour:
\be
Q(G) \sim e^{-0.432 G} \quad (G \to +\infty); \qquad Q(G) \sim e^{-5.058 |G|} \quad (G \to -\infty),
\ee
showing again the strong asymmetry in the profit and loss distribution. 

\begin{figure}
\begin{center}
\psfig{file=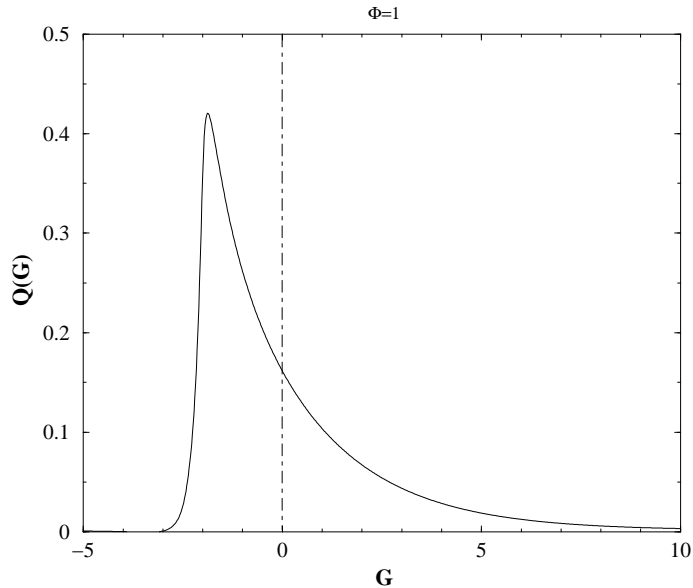,width=8cm,angle=270} 
\end{center}
\caption{Shape of the distribution of gains, $Q(G)$, for a rescaled channel width $\Phi=1$. Note the strong 
asymmetry of the distribution, which peaks at $G^* \approx -1.88$, with a total probability of loss of $0.635$.
When the strategy makes money, the average gain is ${\cal G}=2.164$.}
\label{Fig1}
\end{figure}

The large $\lambda$ behaviour of $\Psi_\lambda(\phi)$ is important to control, in particular to determine accurately
the numerical Fourier transform that gives $Q(G)$. Using Eq. (\ref{ZZ}), we find:
\be
\Psi_\lambda(\Phi) \sim \Re \left[\exp\left(2 i \lambda \Phi - \frac{4}{3} \Phi^{3/2} \sqrt{|\lambda|}\right)\right].
\ee

In the limit $\Phi \to 0$, the distribution becomes maximally skewed. Since the sell threshold is so close to the buy 
threshold, most events correspond to a small ($O(2\Phi)$) immediate loss. Only with a small probability, also of order 
$\Phi$, is the strategy leading to an order $1$ profit. In the small $\Phi$ limit, one finds that the small $\lambda$
expansion of $\Psi_\lambda(\phi)$ reads:
\be\label{large}
\ln \Psi_\lambda(\Phi) \sim -\Phi \left(\sqrt{\pi} \lambda^2 - 2 i \lambda - 2.38..\lambda^4 + ...\right),
\ee
which translates into a diverging skewness, given by 
$\langle G^3 \rangle/\langle G^2 \rangle^{3/2} \approx 1.798/\sqrt{\Phi}$ and a diverging kurtosis 
$\langle G^4 \rangle/\langle G^2 \rangle^{2} \approx 4.545/{\Phi}$. In that limit, $Q(G)$ becomes a $\delta$ peak
at $G=-2 \Phi$ of width $\Phi$ and weight $1 - \Phi$, plus a regular function of total weight $\Phi$. The 
distribution $Q(G)$ decaying exponentially for $G \gg 1$, with a rate $\lambda_+(\Phi \to 0) \approx 0.810 i$, whereas 
for the $G$ negative region, we find that $\lambda_-(\Phi \to 0) \sim  -i/\Phi$, in agreement with our statement that
$Q(G)$ becomes sharply peaked around $G=-2 \Phi$. 

\begin{figure}
\begin{center}
\psfig{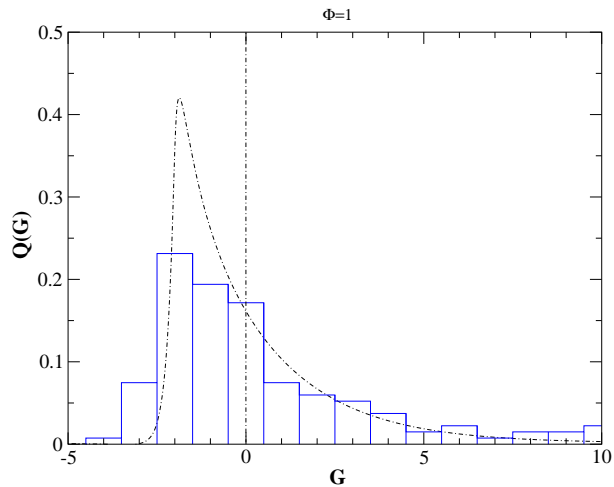} 
\end{center}
\caption{Simulated distribution of gains for a trend following strategy on the Swiss Franc/Dollar, compared to our
theoretical prediction based on a Gaussian model for the returns. As expected, the empirical distribution is 
indeed asymmetric, but also fatter than predicted.}
\label{Fig2}
\end{figure}

We have performed a numerical simulation of the above simple trend following strategy on the Swiss Franc against
Dollar, using 5358 days between 1985 and 2005, with $\tau=5$ days, and choosing the channel width $\Phi=\sigma^2 \tau$,
where $\sigma$ is the historical volatility over the whole time period. The result for $Q(G)$ is given in Fig. 2, and
compared with the theoretical prediction. The agreement is only qualitative, mostly due to the fact 
that the trading is in discrete time (daily) and to non Gaussian character of the returns 
which makes the distribution $Q(G)$ fatter than predicted by the above model (see Fig. 3). 
However, the 
strong asymmetry is indeed observed; in particular, the loss probability is found to be 
$\approx 0.605$, not far from the
theoretical prediction of $0.635$.

\begin{figure}
\begin{center}
\psfig{file=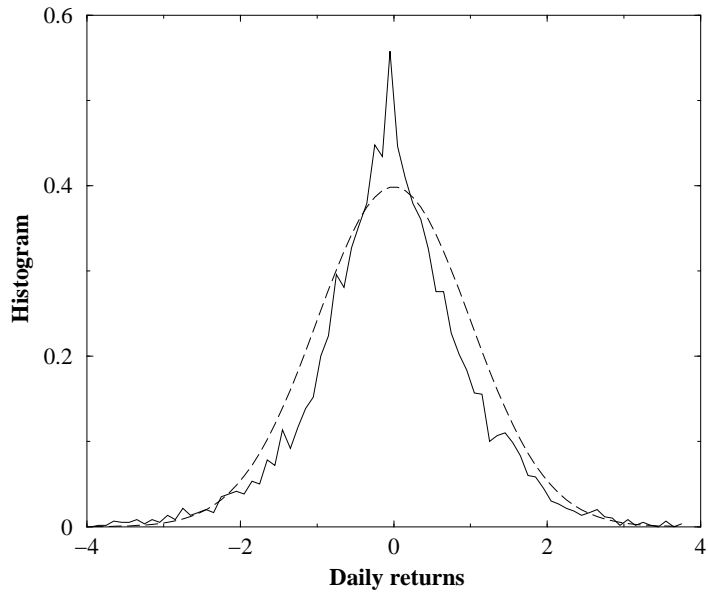,width=8cm,angle=270} 
\end{center}
\caption{Simulated distribution of the daily returns (in units of the daily volatility) of the trend following strategy. 
This distribution is nearly symmetric, but clearly displays non Gaussian tails (a Gaussian distribution is shown in
dotted line).}
\label{Fig3}
\end{figure}

{\it Conclusion}. We have therefore solved exactly a simple model for the profit and losses of a trend following 
strategy, and obtained the analytical shape of the profit distribution per trade. This distribution 
turns out to be
highly non trivial and, most importantly, asymmetric, resembling the distribution 
of an option pay-off. The degree of asymmetry depends continuously on the parameters of 
the strategy and on the volatility of the traded asset; while the average gain per trade is 
always exactly zero, 
the fraction of winning trades decreases from $f=1/2$ for small volatility to $f=0$ for high 
volatility, showing that 
this probability does not give any information on the reliability of the strategy but is indicative of the trading 
style. In fact, we could repeat the same calculations as above for a `mean-reverting' strategy, where the position of
the trade is to sell when the trend indicator is high, and vice-versa. It is clear that the distribution of gains in that
case is the mirror image of that computed above; for a mean reverting strategy, gains are more frequent than losses,
but of a smaller amplitude. Note that the non trivial structure of the gain distribution entirely 
comes from the conditioning on being associated to a {\it given} trade. If one asks, for example, 
for the unconditional 
distribution of the daily returns of the strategy, then it is perfectly symmetrical 
and reproduces exactly the 
return distribution of the underlying asset (see Fig. 3)!

\newpage

\section*{Appendix: Duration of the trades}

In this appendix, we give an alternative derivation of the gain distribution $Q(G)$ which also allows to gather some 
information on their duration. The method presented in the main text is elegant precisely because it gets rids of all
temporal aspects. Suppose that at $t=0$ a buy trade is opened, with $\phi=+\Phi$. We will now focus on $g'$, the profit
accumulated up to time $t$. Let us introduce the quantity $R(\phi,g',t)$ as the probability that the trade is still 
open at time $t$, has accumulated a profit $g'$ and such that the trend indicator is $\phi$. After Fourier transforming on
$g'$, this quantity admits the following path integral representation:
\be
\tilde R(\phi,\lambda,t)=e^{-\lambda^2/2 + (\Phi - i \lambda)^2/2 - (\phi - i \lambda)^2/2-t/\tau} 
\int_{\varphi(t=0)=\Phi - i \lambda}^{\varphi(t)=\phi - i \lambda} {\cal D}\varphi(u) \exp\left[-\frac{1}{2 \sigma^2}
\int_0^t {\rm d}u \left((\frac{d\varphi}{du})^2 + \frac{\varphi^2}{\tau^2} + \lambda^2 \sigma^4 + V(\varphi)\right)\right],
\ee
where $V(\varphi)$ enforces the constraint that $\phi$ never touched the lower edge of the channel $-\Phi$, i.e.
$V(\varphi) = 0$ if $\varphi > -\Phi$ and $V(\varphi) = +\infty $ if $\varphi < -\Phi$. Using standard techniques, 
one sees that the path integral is the Feynman-Kac representation of the imaginary time Green function of the 
quantum harmonic oscillator with an impenetrable wall at $\phi=-\Phi$. Setting again $\sigma^2 \tau=1$ and 
using a wave function representation, one can therefore write:
\be
\tilde R(\phi,\lambda,t)=e^{-\lambda^2/2 + (\Phi - i \lambda)^2/2 - (\phi - i \lambda)^2/2}
\sum_m \psi_m(\phi - i \lambda) \psi_m(\Phi - i \lambda) e^{-E_m t/\tau},
\ee
where $\psi_m$ and $E_m$ are the eigenvectors and eigenvalues of a quantum harmonic oscillator, obeying:
\be
\left[-\frac12 \frac{\partial^2}{\partial \phi^2} +\frac{\phi^2}{2}+\frac{\lambda^2-1}{2}\right]\psi_m(\phi) = E_m \psi_m(\phi),
\ee
with the following boundary conditions: $\psi_m(-\Phi)=0$ (hard wall condition) and $\psi_m(\phi \to \infty) \to 0$.
These two conditions lead to a quantized spectrum of eigenvalues, indexed by an integer number $m$; as expected,
the $\psi_m(\phi)$ can again be written in terms of Weber functions \cite{HarmOsc}. 

From $\tilde R(\phi,\lambda,t)$ one can compute the flux of fictitious particles just hitting the wall at time $t=T$ 
and leaving the system, given by:
\be
J(\phi,\lambda,t=T)= -\frac{1}{2\tau} \left.\frac{\partial}{\partial \phi} 
\tilde R(\phi,\lambda,t) \right|_{\phi=-\Phi},
\ee
which, for $\phi=\Phi$ is precisely the joint probability that the profit of the trade is $G$ and its duration is $T$ 
(Fourier transformed over $G$.) The integral over all $G$'s, corresponding to $\lambda=0$, gives the
unconditional distribution of trade times. The result can be written in a fully explicit way 
if the trade closing condition is at $\phi=0$, in which case the eigenvectors $\psi_m$ are simply 
the odd levels of the harmonic oscillator and can be expressed in terms of Hermite polynomials. 
More generally, the distribution of duration decays at large times as 
$\exp(-E_0 T/\tau)$, where $E_0$ is the ground state energy of the constrained Harmonic oscillator. 
One can also check that the integral over all $T$s of $J(\phi=-\Phi,\lambda,T)$ obeys the same 
{\sc ode} (with respect to the initial condition $\phi=+\Phi$ and up to a sign change of $\lambda$) as $\Psi_\lambda$ in the main text 
, as it should since the former quantity then becomes 
the Fourier transform of $Q(G)$, with the same boundary condition ($g' \equiv 0$ when $\phi=-\Phi$).


\begin{thebibliography}{99}

\bibitem{Book} see e.g. J.P. Bouchaud, M. Potters, {\it Theory of Financial Risks and Derivative Pricing}, 
Cambridge University Press (2004).

\bibitem{Grad} see e.g. I. Gradshteyn, I. Ryzhik, {\it Tables of Integrals, Series and Products}, Academic Press, (1980) 
p. 1057-1059

\bibitem{HarmOsc} see: W. N. Mei, Y. C. Lee, {\it Harmonic Oscillator with potential barriers}, J. Phys. A {\bf 16}, 1623 
(1983) 


\end{thebibliography}
\end{document}